\documentclass[12pt]{article}
 \pdfoutput=1
\usepackage{amssymb}
\usepackage{amsmath}
\usepackage{amstext}
\usepackage{graphicx,epsfig}
\usepackage{epsfig}
\usepackage{verbatim} 
\usepackage{fancybox}
\usepackage{color}
\usepackage{ulem}
\usepackage{enumitem}
\usepackage{subfigure}
\usepackage{bbm}
\usepackage{parskip}
\usepackage{cite}

\newcommand{\Comment}[1]{{}}
\definecolor{MyDarkBlue}{rgb}{0.15,0.15,0.45}
\usepackage[linktocpage=true]{hyperref}
\hypersetup{
colorlinks=true,
citecolor=MyDarkBlue,
linkcolor=MyDarkBlue,
urlcolor=MyDarkBlue,
pdfauthor={Kurt Hinterbichler},
pdftitle={Ghost-Free Derivative Interactions for a Massive Graviton},
pdfsubject={hep-th}
}

\newcommand{\be}{\begin{equation}}
\newcommand{\ee}{\end{equation}}
\newcommand{\bea}{\begin{eqnarray}}
\newcommand{\eea}{\end{eqnarray}}
\newcommand{\beas}{\begin{eqnarray*}}
\newcommand{\eeas}{\end{eqnarray*}}
\newcommand{\nn}{\nonumber}

\def\({\left(}
\def\){\right)}

\numberwithin{equation}{section}

\begin{document}


\begin{center}
{\LARGE { Ghost-Free Derivative Interactions for a  \\ \vspace{.2cm} Massive Graviton}}
\end{center} 
 \vspace{1truecm}
\thispagestyle{empty} \centerline{
{\large  { Kurt Hinterbichler}}\footnote{E-mail: \Comment{\href{mailto:khinterbichler@perimeterinstitute.ca}}{\tt khinterbichler@perimeterinstitute.ca}}
                                                          }

\vspace{1cm}

\centerline{{\it 
Perimeter Institute for Theoretical Physics,}}
 \centerline{{\it 31 Caroline St. N, Waterloo, Ontario, Canada, N2L 2Y5 }} 

\begin{abstract}

If the kinetic interactions of a Lorentz-invariant massive graviton are Einstein- Hilbert, then the only possible potential terms free from the Boulware-Deser ghost are those of de Rham, Gabadadze and Tolley (dRGT). We point out there are other possibilities if the kinetic terms are not required to be Einstein-Hilbert. We construct pseudo-linear ghost-free potentials, terms that derive in a natural way from the lin- ear theory. The simplicity of this approach allows us to construct diffeomorphism non-invariant higher-derivative interaction terms that do not introduce ghosts. We conjecture that these terms should have counterparts in the full dRGT theory. These terms would introduce new free parameters into the theory and may change some of the conclusions heretofore drawn.

\end{abstract}

\newpage

\tableofcontents
\newpage

\section{Introduction}
\parskip=5pt
\normalsize

Much of the recent interest in massive gravity has stemmed from a solution to the problem of finding interaction terms for a massive graviton that propagate the same number of degrees of freedom as the free theory.  This solution is dRGT theory (de Rham, Gabadadze and Tolley \cite{deRham:2010ik,deRham:2010kj}), which consists of the most general potential terms which may be added to the Einstein-Hilbert term without introducing unwanted degrees of freedom.  We will be interested in the question of whether there are non-potential terms -- terms with derivative interactions -- that are also solutions to this problem, with the goal of characterizing the most general solution.

The free relativistic massive spin-2 particle of mass $m$ in $D$-dimensional Minkowski space ($D\geq 3$) is described by a symmetric tensor field $h_{\mu\nu}$ governed by the Lagrangian of Fierz and Pauli \cite{Fierz:1939ix},
\be{\cal L}=-\frac{1}{2}\partial_\lambda h_{\mu\nu}\partial^\lambda h^{\mu\nu}+\partial_\mu h_{\nu\lambda}\partial^\nu h^{\mu\lambda}-\partial_\mu h^{\mu\nu}\partial_\nu h+\frac{1}{2}\partial_\lambda h\partial^\lambda h-\frac{1}{2}m^2(h_{\mu\nu}h^{\mu\nu}-h^2). \label{fp}\ee
This action propagates precisely the $D(D-1)/2-1$ degrees of freedom of a massive graviton (five of them in $D=4$).

First consider the case $m=0$.  The action \eqref{fp} develops a linear diffeomorphism invariance,
\be \delta h_{\mu\nu}=\partial_\mu\xi_\nu+\partial_\nu\xi_\mu,\label{lindiffeomorphism}\ee
with one-form gauge parameter $\xi_\mu(x)$, and describes the $(D-1)(D-2)/2-1$ helicity states of a massless graviton (two of them in $D=4$).  In this case, the only two-derivative interaction terms that may be added which preserve the number of degrees of freedom of the free theory are those of the Einstein-Hilbert term $\sim \sqrt{-g}R$ \cite{Gupta:1954zz,Kraichnan:1955zz,Weinberg:1965rz,Deser:1969wk,Boulware:1974sr,Fang:1978rc,Wald:1986bj}.  We may think of the Einstein-Hilbert term as the non-linear counterpart of \eqref{fp} with $m=0$.  

If we allow terms with no derivatives, the only additional possibility is a cosmological constant $\sim \sqrt{-g}$.  

If we allow for terms with $d>2$ derivatives, then the Lanczos-Lovelock terms may be added
\cite{Lanczos:1938sf,Lovelock:1971yv}
\be {\cal L}^{(d)}\sim \sqrt{-g} g^{\mu_1\nu_1\cdots \mu_d\nu_d}R_{\mu_1\mu_2\nu_1\nu_2}R_{\mu_3\mu_4\nu_3\nu_4}\cdots R_{\mu_{d-1}\mu_d\nu_{d-1}\nu_d},\label{lovelock}\ee
where $g^{\mu_1\nu_1\cdots \mu_d\nu_d}$ is defined as the product $g^{\mu_1\nu_1}\cdots g^{\mu_d\nu_d}$ anti-symmetrized over the $\nu$ indices.  ${\cal L}^{(0)}\sim \sqrt{-g}$ is the cosmological term,  ${\cal L}^{(2)}\sim \sqrt{-g}R$ is the Einstein-Hilbert term, ${\cal L}^{(4)}\sim \sqrt{-g}\( R^2 - 4 R_{\mu\nu} R^{\mu\nu}
+ R_{\mu\nu\rho\sigma} R^{\mu\nu\rho\sigma} \)$ is the Gauss-Bonnet term, and so on.  ${\cal L}^{(d)}$ is a total derivative for $d=D$ and vanishes identically for $d>D$, so there are non-trivial higher-derivative possibilities only when $D>4$.

Expanding around flat space, $g_{\mu\nu}=\eta_{\mu\nu}+h_{\mu\nu}$, the expansion of the Lovelock term ${\cal L}^{(d)}$ starts at order $d/2$ in powers of $h_{\mu\nu}$ \cite{Zumino:1985dp}, 
\be {\cal L}^{(d)}=a_{d/2}+a_{d/2+1}+\cdots.\ee
The first term, $a_{d/2}$, is a total derivative,
 \bea \label{totald0} a_{d/2}&\sim& \eta^{\mu_1\nu_1\cdots \mu_d\nu_d}\partial_{\mu_1}  \partial_{\nu_1}  h_{\mu_2\nu_2}\partial_{\mu_3}  \partial_{\nu_3}  h_{\mu_4\nu_4}\cdots \partial_{\mu_{d-1}}  \partial_{\nu_{d-1}}  h_{\mu_d\nu_d} \nn\\
 &\sim& \partial_{\mu_1} \(\eta^{\mu_1\nu_1\cdots \mu_d\nu_d} \partial_{\nu_1}  h_{\mu_2\nu_2}\partial_{\mu_3}  \partial_{\nu_3}  h_{\mu_4\nu_4}\cdots \partial_{\mu_{d-1}}  \partial_{\nu_{d-1}}  h_{\mu_d\nu_d}\),
 \eea
where $\eta^{\mu_1\nu_1\cdots \mu_d\nu_d}$ is now the product of flat metrics $\eta^{\mu_1\nu_1}\cdots \eta^{\mu_d\nu_d}$ anti-symmetrized over the $\nu$ indices.
The first non-trivial term is $a_{d/2+1}$ \cite{Cnockaert:2005jw}, 
\be \label{quaslin} a_{d/2+1}\sim \eta^{\mu_1\nu_1\cdots \mu_{d+1}\nu_{d+1}}\partial_{\mu_1}\partial_{\nu_1}h_{\mu_2\nu_2}\partial_{\mu_3}  \partial_{\nu_3}  h_{\mu_4\nu_4}\cdots\partial_{\mu_{d-1}}\partial_{\nu_{d-1}}h_{\mu_d\nu_d}\, h_{\mu_{d+1}\nu_{d+1}}.\ee
Since this is the first non-trivial term in the expansion of a diffeomorphism invariant around flat space, it is invariant (up to a total derivative) under linearized diffeomorphisms \eqref{lindiffeomorphism}, as can be explicitly checked by using the anti-symmetry of the $\eta$ symbol and integrations by parts.  The terms \eqref{quaslin} yield second order equations of motion for $h$, just as their fully non-linear counterparts, the Lovelock terms, yield second order equations of motion for the metric\footnote{See \cite{Deser:2011zk} for a detailed analysis of the $d=4$ case.}.  One might call these ``pseudo-linear'' terms -- they are non-linear in $h$, but they are the leading term in the expansion in small fluctuations of a fully diffeomorphism invariant quantity, and are invariant under the gauge symmetries of the linear theory.

This opens up a new possibility for interacting gravity: we may add the terms \eqref{quaslin} to the free graviton action.  This gives an interacting theory (only when $D>4$, otherwise there are no non-trivial terms except the zero-derivative tadpole term $\sim h$), which preserves the number of degrees of freedom of the free theory, but unlike Einstein-Hilbert, does not modify the linear gauge symmetry \eqref{lindiffeomorphism} at higher order.

In this paper, we develop an analogous story for the massive graviton.  We seek to add interaction terms to \eqref{fp} with $m\not=0$ such that the number of degrees of freedom is unchanged from the free case.  Diffeomorphism invariance is broken, so naively any interaction term is allowed, but most choices will lead to a theory with an extra degree of freedom, the Boulware-Deser ghost \cite{Boulware:1973my}.  If the two-derivative non-linearities are chosen to be those of Einstein-Hilbert, and no higher derivative interactions are considered, then one only has to choose the zero-derivative interactions.  The only possibilities that do not introduce the Boulware-Deser ghost  \cite{Hassan:2011hr,Hassan:2011ea} are those of de Rham, Gabadadze and Tolley\footnote{As pointed out in \cite{Deser:2012qx}, for different reasons some of these terms were written down much earlier \cite{Zumino}.  However, since \cite{Zumino} precedes \cite{Boulware:1973my}, none of the issues regarding the absence or presence of ghosts could be appreciated at the time.} (dRGT theory) \cite{deRham:2010ik,deRham:2010kj} (see \cite{Hinterbichler:2011tt} for a review).  The dRGT terms are
\be {\cal V}_n\sim \sqrt{-g}g^{\mu_1\nu_1\cdots \mu_d\nu_d}{\cal K}_{\mu_1\nu_1}{\cal K}_{\mu_2\nu_2}\cdots{\cal K}_{\mu_n\nu_n}, \ \ \ 0\leq n\leq D.\label{drgtterms}\ee
where ${\cal K}^\mu_{\ \nu}\equiv\delta^{\mu}_{\ \nu}-\(\sqrt{g^{-1}\eta}\)^\mu_{\ \ \nu}$.  Of these $D+1$ terms, one combination is a cosmological constant $\sim \sqrt{-g}$, one combination is a literal constant $\sim \sqrt{-\eta}$, one further combination can be identified with the graviton mass, and the remaining $D-2$ constants represent genuine new parameters of the theory.  The terms \eqref{drgtterms} are to the mass terms of \eqref{fp} what Einstein-Hilbert is to the kinetic terms of \eqref{fp}.

In what follows, we will consider ghost free potentials that are analogous to the pseudo-linear terms \eqref{quaslin}.  As we will show, if one keeps the kinetic terms of the linear theory \eqref{fp} unchanged, then the following potential terms may be added without introducing new degrees of freedom,
\be \sim \eta^{\mu_1\nu_1\cdots \mu_n\nu_n}h_{\mu_1\nu_1}\cdots h_{\mu_n\nu_n},\ \ \ \ 0\leq n\leq D.\label{zerod0}\ee
These are pseudo-linear mass terms, and they are to the dRGT mass terms \eqref{drgtterms} what the pseudo-linear terms \eqref{quaslin} are to the Lovelock terms \eqref{lovelock}.  Indeed, they are the leading terms in the expansions of \eqref{drgtterms} around flat space.

The relative simplicity of the pseudo-linear approach will allow us to write down all the higher derivative terms which are ghost-free.  In particular, there are two-derivative interaction terms, cubic order and higher in $h$, which, like the potential terms \eqref{zerod0}, are are not diffeomorphism invariant but nevertheless have the proper constraint structure so that new degrees of freedom are not introduced.  Analogous terms with $\geq 4$ derivatives also exist.  We will label these terms ${\cal L}_{d,n}$, where $d$ denotes the number of derivatives and $n$ the number of $h$'s.  The terms ${\cal L}_{0,n}$, $0\leq n\leq D$ are the potential terms \eqref{zerod0} (in reality there are only $D$ such terms, since ${\cal L}_{0,0}$ is just a constant).
The linear kinetic term is ${\cal L}_{2,2}$.  In general, there are terms ${\cal L}_{d,n}$ for $d/2\leq n\leq D-d/2$, for a total of $D-d+1$ terms with $d$-derivatives.  The lowest term ${\cal L}_{d,d/2}$ is the total derivative \eqref{totald0}, so there are in fact only $D-d$ terms.  The lowest of these, ${\cal L}_{d,d/2+1}$, is the pseudo-linear Lovelock term \eqref{quaslin}, invariant under the linearized diffeomorphisms \eqref{lindiffeomorphism}.  The remaining $D-d-1$ terms are not invariant under linearized diffeomorphisms and are the genuinely new derivative interactions of a massive graviton.

In $D=3$, there are no novel derivative interactions.  In $D=4$ there is a single new term: a cubic interaction with two derivatives,
\be \label{quaslin4d} {\cal L}_{2,3}\sim \eta^{\mu_1\nu_1\cdots \mu_4\nu_4}\partial_{\mu_1}\partial_{\nu_1}h_{\mu_2\nu_2}\, h_{\mu_3\nu_3}h_{\mu_4\nu_4}.\ee
This is the same as the non-Einstein two-derivative term arrived at in the investigations of \cite{Folkerts:2011ev}.
To get terms with four or more derivatives, we must go to $D>4$.  

In the massless case, there was a one-to-one correspondence between the possible fully non-linear terms -- the Lovelock terms \eqref{lovelock} -- and their pseudo-linear counterparts \eqref{quaslin}, the pseudo-linear term being the leading term in the expansion of the corresponding fully non-linear term around flat space.  We conjecture that the same is true in the massive case.  This conjecture is true for the zero derivative interactions; the dRGT mass terms \eqref{drgtterms} have their pseudo-linear counterparts in \eqref{zerod0}.  If it is true for the higher derivative terms as well, then fully non-linear terms corresponding to ${\cal L}_{d,n}$ for $d\geq 2$ also exist.   These are as yet unknown (other than ${\cal L}_{d,d/2+1}$, which correspond to the Lovelock terms \eqref{lovelock}).  Thus if our conjecture is correct, there should exist new diffeomorphism non-invariant ghost-free terms for the fully non-linear massive gravity.  In particular, dRGT theory in $D=4$ as presently studied is incomplete; there should be a two-derivative, diffeomorphism non-invariant, ghost free term corresponding to \eqref{quaslin4d}.  This term may be added, providing another parameter in the theory, and possibly changing some of the conclusions (e.g. \cite{Gumrukcuoglu:2011zh,Deser:2012qx}) which have been drawn from the theory containing only the zero-derivative terms \eqref{drgtterms}.

The analysis of the pseudo-linear terms is much simpler than the analysis of their fully non-linear counterparts.  In particular, the {\it linear} St\"uckelberg replacement $h_{\mu\nu}\rightarrow h_{\mu\nu}+\partial_\mu A_\nu+\partial_\nu A_\mu+2\,\partial_\mu\partial_\nu\phi$ is all that is needed to easily extract the complete dynamics of the helicity components.  We will be able to easily derive the decoupling limit Lagrangians for all the various interactions ${\cal L}_{d,n}$, including the contributions from the vector modes.  

The scalar-tensor sector of the decoupling limit of the pseudo-linear potential terms \eqref{totald0} turns out to be exactly the same as that of their fully non-linear dRGT counterparts.  This sector of the decoupling limit is blind to the difference between the pseudo-linear and fully non-linear terms.  Conjecturing that this remains true for the derivative interactions, we may derive the decoupling limit of the unknown higher-derivative terms.  In particular, the missing 2 derivative term in $D=4$ can make a contribution to the decoupling limit, suppressed by the same scale as the known terms.  This new contribution involves two powers of $h$, one power of $\partial\partial\phi$, and 2 extra derivatives, 
\be {\cal L}_{2,3}\supset \eta^{\mu_1\nu_1\cdots \mu_4\nu_4}\partial_{\mu_1}\partial_{\nu_1}h_{\mu_2\nu_2}\, h_{\mu_3\nu_3}\partial_{\mu_4}\partial_{\nu_4}\phi.
\ee
Analysis of solar-system constraints, stability of matter sources, and other calculations that rely only on the decoupling limit can then be done without knowing the full term.

In what follows, we first describe in Section \ref{ham} a general Hamiltonian analysis which will suffice to show that all of the various terms we introduce are ghost free.  We then introduce the possible terms in Section \ref{termsec}, and their decoupling limits in Section \ref{stuksection}, followed by conclusions and speculations in Section \ref{conc}.  In Appendix \ref{screensec}, we display a simple degravitating solution within the pseudo-linear theory, to illustrate how it shares many of the features of the full theory. 

\section{\label{ham}Hamiltonian analysis}
\parskip=5pt
\normalsize

One way to see that the free massive graviton action \eqref{fp} propagates the proper number of degrees of freedom is to perform a Hamiltonian analysis.  The time derivatives in the Lagrangian can be integrated by parts in such a way that they never appear acting on $h_{0i}$ or $h_{00}$, and never more than once on $h_{ij}$.  Furthermore, the Lagrangian is linear in $h_{00}$, and the part multiplying $h_{00}$ does not depend on $h_{0i}$ or on any time derivatives.  In sum, the Lagrangian takes the form
\be {\cal L}={\cal F}\( h_{ij},\dot h_{ij},h_{0i}\)+h_{00}\,{\cal G}\(h_{ij}\),\label{structure}\ee
where ${\cal F}$ is some function of $h_{ij},\dot h_{ij},h_{0i}$ and ${\cal G}$ is some function of $h_{ij}$ only.  Both ${\cal F}$ and ${\cal G}$  may have arbitrary dependence on spatial derivatives of the various arguments as well, which we suppress, since these do not affect the Hamiltonian analysis.  We will find that the various interaction terms we introduce all conform to the structure \eqref{structure}, with various ${\cal F}$ and ${\cal G}$, so we need only do the Hamiltonian analysis once for generic ${\cal F}$ and ${\cal G}$ and it will apply to all the terms.

We Legendre transform with respect to the spatial components $h_{ij}$ only.  The canonical momenta are found as functions of $h_{ij},\dot h_{ij}$, and $h_{0i}$, 
\be \pi^{kl}\( h_{ij},\dot h_{ij},h_{0i}\)={\partial {\cal L}\over \partial \dot h_{kl}}={\partial {\cal F}\over \partial \dot h_{kl}}. \ee
In all the cases we are interested in, this may be inverted to find the velocities as functions of the momenta, $h_{ij}$ and $h_{0i}$, 
\be \pi^{kl}=\pi^{kl}\( h_{ij},\dot h_{ij},h_{0i}\)\Rightarrow \dot h_{kl}=\dot h_{kl}\(\pi^{ij}, h_{ij},h_{0i}\).\ee

The Hamiltonian takes the form
\be {\cal H}=\pi^{kl}\dot h_{kl}-{\cal L}=\pi^{kl}\dot h_{kl}\(\pi^{ij}, h_{ij},h_{0i}\)-{\cal F}\( h_{ij},\dot h_{ij}\(\pi^{ij}, h_{ij},h_{0i}\),h_{0i}\)-h_{00}\,{\cal G}\(h_{ij}\).\label{hamlate}\ee
The $h_{0i}$ appear algebraically and may be eliminated by their own equations of motion (if instead they appear linearly, then we have the linear diffeomorphism invariance \eqref{lindiffeomorphism} and the $h_{0i}$ enforce the momentum constraints), and the solution does not involve $h_{00}$,
\be {\delta {\cal H}\over \delta h_{0k} }=0\Rightarrow h_{0k}=h_{0k}\( \pi^{ij},h_{ij}\).\ee

Substituting back into \eqref{hamlate}, we see that the action remains linear in $h_{00}$, enforcing a constraint that (along with its secondary constraint) eliminates the Boulware-Deser extra degree of freedom.

\section{\label{termsec}Ghost-free interaction terms}

In this Section we write down the ghost-free pseudo-linear interaction terms discussed in the introduction.
We seek to write down interaction terms that conform to the structure of \eqref{structure}.  We label the terms ${\cal L}_{d,n}$, where $d$ is the number of derivatives in the term, and $n$ is the number of fields.

We make use of the symbol $\eta^{\mu_1\nu_1\mu_2\nu_2\cdots\mu_n\nu_n}$, defined as the product of $\eta$'s anti-symmetrized over one set of indices,
\be
\label{tensor} 
\eta^{\mu_1\nu_1\mu_2\nu_2\cdots\mu_n\nu_n}\equiv{1\over n!}\sum_p\left(-1\right)^{p}\eta^{\mu_1p(\nu_1)}\eta^{\mu_2p(\nu_2)}\cdots\eta^{\mu_np(\nu_n)} \ ,
\ee 
where the sum is over all permutations of the $\nu$ indices, with $(-1)^p$ the sign of the permutation.  The tensor~(\ref{tensor}) is anti-symmetric in the $\mu$ indices, anti-symmetric the $\nu$ indices, and hence symmetric under interchange of any $\mu,\nu$ pair with any other.
(We may also write it as a product of two epsilon symbols, $\eta^{\mu_1\nu_1\cdots \mu_n\nu_n}\sim \epsilon^{\mu_1\ldots\mu_n\,\sigma_{n+1}\ldots\sigma_D}\epsilon^{\nu_1\ldots\nu_n}_{\ \ \ \ \ \ \ \sigma_{n+1}\ldots\sigma_D}$, making the anti-symmetries manifest.)  This anti-symmetry will allow us to ensure the structure \eqref{structure}.

\subsection{Zero-derivative terms}

We start with the zero-derivative terms.  There are $D+1$ zero-derivative terms, which are the possible contractions of powers of $h$ with the $\eta$ symbol \eqref{tensor},
\be {\cal L}_{0,n}\sim\eta^{\mu_1\nu_1\cdots \mu_n\nu_n}h_{\mu_1\nu_1}\cdots h_{\mu_n\nu_n},\ \ \ \ 0\leq n\leq D.\label{zerod}\ee
These are nothing but the symmetric polynomials of $h_{\mu\nu}$,
\bea {\cal L}_{0,0}&\sim&1 \, ,\nn\\
{\cal L}_{0,1}&\sim&[h] \, ,\nn\\
{\cal L}_{0,2}&\sim&\left. [h]^2-[h^2]\right.\, ,\nn\\
{\cal L}_{0,3}&\sim&\left. [h]^3-3[h][h^2]+2[h^3]\right. \, ,\nn\\
&&\vdots \nn\\
{\cal L}_{0,D}&\sim&\det h,
\eea
where the square brackets denote traces of powers of the matrix $h^\mu_{\ \nu}$.

Note that ${\cal L}_{0,0}$ is not really a term because it is a constant, so in fact we have only $D$ genuine two-derivative terms.   The lowest order term, the tadpole ${\cal L}_{0,1}\sim h$, is the linearization of $\sqrt{-g}$ and hence is invariant under linearized diffeomorphisms \eqref{lindiffeomorphism}. The $n=2$ term is the Fierz-Pauli mass term ${\cal L}_{0,2}\sim h_{\mu\nu}h^{\mu\nu}-h^2$.  The higher terms are the pseudo-linear counterparts of the dRGT mass terms \eqref{drgtterms}.

Using the anti-symmetries of the $\eta$ symbol \eqref{tensor} it is easy to see that the terms \eqref{zerod} take the form \eqref{structure}, and hence are ghost free: if any one of the $h$'s in \eqref{zerod} carries the indices $00$, then no other $h$ can carry a $0$ index, so the terms are linear in $h_{00}$, and $h_{00}$ multiplies a function of $h_{ij}$ only.  Only the structure of contractions in \eqref{zerod} can produce this property.

\subsection{\label{2dterms}Two-derivative terms}

We now generalize the structure of the terms \eqref{zerod} to include two derivatives.  The only way to insert derivatives so as to give the right Hamiltonian structure \eqref{structure} is through the following $D-1$ two derivative terms, 
\be {\cal L}_{2,n}\sim \eta^{\mu_1\nu_1\cdots \mu_{n+1}\nu_{n+1}}\partial_{\mu_1}\partial_{\nu_1}h_{\mu_2\nu_2}\, h_{\mu_3\nu_3}\cdots h_{\mu_{n+1}\nu_{n+1}},\ \ \ \ 1\leq n\leq D-1.\label{2d}\ee
Again we can see using the anti-symmetry of the $\eta$ symbol \eqref{tensor} how these terms take the form \eqref{structure}: if any one of the $h$'s in \eqref{zerod} carries the indices $00$, then no other $h$ or derivative can carry a $0$ index, so the term is linear in $h_{00}$, and $h_{00}$ multiplies a function of $h_{ij}$ only.  If, in the $\partial\partial h$ factor, one of the derivatives and one of the $h$ indices carry $0$'s (leading to a potentially dangerous $\dot h_{0i}$), then no other $h$ or derivative carries a $0$, and we may integrate by parts the $\partial_0$ off of $h_{0i}$.  If both derivatives carry a $0$ (leading to a worrisome double time derivative) then all the remaining indices, those of the $h$'s, must take spatial values, and we may integrate by parts one of the time derivatives, resulting in terms with only first time derivatives of the spatial components, $\dot h_{ij}$.

Note that ${\cal L}_{2,1}$ is not in fact a term because it is a total derivative (it is the linearization of $\sim\sqrt{-g}R$) so in fact we have only $D-2$ genuine two-derivative terms.  The term ${\cal L}_{2,2}$ is nothing but the kinetic term of \eqref{fp}, i.e. the quadratic part of $\sim\sqrt{-g}R$, and so it is invariant under linearized diffeomorphisms \eqref{lindiffeomorphism}.  

The higher terms ${\cal L}_{2,n}$ for $n>2$ are not diffeomorphism invariant.  These pseudo-linear terms should correspond to fully non-linear, non-diffeomorphism invariant two-derivative terms (as yet unknown) which may be added to massive gravity.

In $D=4$, there is one such novel two-derivative term, cubic in $h$,
\be\label{miss4} {\cal L}_{2,3}\sim \eta^{\mu_1\nu_1\cdots \mu_4\nu_4}\partial_{\mu_1}\partial_{\nu_1}h_{\mu_2\nu_2}\, h_{\mu_3\nu_3}h_{\mu_4\nu_4}.\ee
Thus we conjecture that there is one missing term in dRGT theory in $D=4$, as it is presently studied.  This term should be a non-diffeomorphism invariant term with two derivatives whose lowest order expansion around flat space produces \eqref{miss4}.

\subsection{$d$-derivative terms}

The pattern can be continued to an arbitrary numbers of derivatives, $d\geq 4$.  There will be $D-d+1$ terms with $d$-derivatives,
\bea && {\cal L}_{d,n}\sim \eta^{\mu_1\nu_1\cdots \mu_{n+d/2}\nu_{n+d/2}}\partial_{\mu_1}\partial_{\nu_1}h_{\mu_2\nu_2}\cdots\partial_{\mu_{d-1}}\partial_{\nu_{d-1}}h_{\mu_d\nu_d}\, h_{\mu_{d+1}\nu_{d+1}}\cdots h_{\mu_{n+d/2}\nu_{n+d/2}}, \nn\\ && \ \ \ \ d/2\leq n\leq D-d/2.\label{dd}\eea

The lowest term, $ {\cal L}_{d,d/2}$, is the total derivative \eqref{totald0} -- the first term in the expansion of the $d$-the order Lovelock invariant \eqref{lovelock} around flat space -- so we have only $D-d$ genuine $d$-derivative terms.  The first non-trivial term term, ${\cal L}_{d,d/2+1}$, is the same as \eqref{quaslin}, the leading non-trivial term in the expansion of the $d$-th order Lovelock, and is invariant under linearized diffeomorphisms \eqref{lindiffeomorphism}.  The remaining terms ${\cal L}_{d,d/2+1}$, for $n\geq d/2+2$, do not have this symmetry.  

These higher derivative pseudo-linear terms are all ghost-free.  They produce second order equations of motion for $h$, and using the anti-symmetry of the $\eta$ symbol and suitable integrations by parts of time derivatives (as described for the two derivative terms in Section \ref{2dterms}), it is again straightforward to see that they conform to the structure of \eqref{structure}.

These terms with four or more derivatives only come in for spacetime dimensions $D>4$.  They should correspond to fully non-linear ghost-free higher derivative interactions which may be added to dRGT theory in higher dimensions.

\section{\label{stuksection}St\"uckelberg and decoupling limits}

In this section we study the effective field theory of the pseudo-linear terms and derive their decoupling limit, the limit where the graviton mass goes to zero, the Planck mass (i.e. the inverse of the non-linear interaction strength) goes to infinity, and the leading strong coupling scale is held fixed. 
Studying the theory in the decoupling limit through the St\"uckelberg trick is a clean way to see the non-linear dynamics of the various helicity components of the massive graviton in the high-energy regime \cite{ArkaniHamed:2002sp,Creminelli:2005qk,deRham:2010kj}.  

The St\"uckelberg analysis of the pseudo-linear theory is much easier than in the fully non-linear dRGT theory.  This is because we need only restore {\it linear} gauge invariance, so we introduce fields $A_\mu$ and $\phi$ governing the helicity one and helicity zero longitudinal modes through a replacement patterned after the linear gauge symmetry \eqref{lindiffeomorphism}, followed by a Maxwell $U(1)$ symmetry,
\be h_{\mu\nu}\rightarrow h_{\mu\nu}+\partial_\mu A_\nu+\partial_\nu A_\mu+2\,\partial_\mu\partial_\nu\phi.\label{stukerep}\ee
After this replacement, we have two gauge symmetries, a linear diffeomorphism with gauge parameter $\xi_\mu(x)$ and a Maxwell $U(1)$ with gauge parameter $\lambda(x)$, 
\bea \label{stuksym} \delta h_{\mu\nu}&=&\partial_\mu \xi_\nu+\partial_\nu \xi_\mu,\ \ \delta A_\mu=\partial_\mu\lambda-\xi_\mu, \ \ \delta\phi=-\lambda. \eea

To determine the scales in the effective theory, we must specify the scales with which all the various interaction terms are to be introduced. 
In the standard dRGT theory in $D=4$, the potential term is chosen to carry an overall factor of $m^2 M_P^2$, so that it is suppressed relative to the Einstein-Hilbert term by two powers of $m/\partial$.  The smallest strong coupling scale then turns out to be $\Lambda_3\sim (m^2M_P)^{1/3}$.  

We will follow this scaling with the pseudo-linear terms.  Going back to general $D$, the kinetic term ${\cal L}_{2,2}$ will carry a power of $M_P^{D-2}$, and the mass terms ${\cal L}_{0,n}$ will carry a power of $M_P^{D-2}m^2$.  With this, the canonically normalized fields are determined to be
\be \hat h\sim M_P^{D/2-1} h,\ \ \  \hat A\sim M_P^{D/2-1} m A,\ \ \  \hat \phi\sim M_P^{D/2-1} m^2\phi,\ee
and the leading strong coupling terms will be interactions between a single tensor and $n-1$ scalars, carrying the scale
\be\label{scale}\Lambda_{D+2\over D-2}\sim \(m^{4\over D-2}M_P\)^{D-2\over D+2} .\ee

For the derivative interactions ${\cal L}_{d,n}$, their scaling will be determined by two requirements: that the term not lower the cutoff by introducing strong interactions at a scale lower than \eqref{scale}, and that it contribute new non-trivial terms at the scale \eqref{scale} (with the exception of the diffeomorphism invariant terms ${\cal L}_{d,d/2+1}$, as these will not contribute any St\"uckelberg fields).  This fixes the scaling to be $\sim M_P^{D-2}m^{2-d}{\cal L}_{d,n}$, that is, each new derivative gets suppressed by a power of $m$.

The decoupling limit is
\be m\rightarrow 0,\ \ \ M_P\rightarrow \infty,\ \ \ \Lambda_{D+2\over D-2}\ {\rm fixed}.\ee
In this limit, the St\"uckelberg gauge symmetry \eqref{stuksym} reduces to linearized diffeomorphisms \eqref{lindiffeomorphism} acting only on the helicity 2 field, a Maxwell $U(1)$ acting only on $A_\mu$, and the longitudinal scalar mode becomes gauge invariant,
\bea \label{decsym} \delta h_{\mu\nu}&=&\partial_\mu \xi_\nu+\partial_\nu \xi_\mu,\ \ \delta A_\mu=\partial_\mu\lambda,\  \ \delta\phi=0. \eea 

In this section, we derive the decoupling limits of the various pseudo-linear terms introduced in Section \ref{termsec}.
Since the St\"uckelberg symmetry is linear, terms of various orders do not mix, and the analysis of the decoupling limit becomes simple.
As we will see, the decoupling limit in the scalar-tensor sector turns out to be exactly the same for the pseudo-linear theory as it is for the fully non-linear theory.  

\subsection{Zero-derivative terms}

When the replacement \eqref{stukerep} is made in the zero-derivative terms \eqref{zerod}, there are no scalar self-interaction terms because they become a total derivative,
\bea {\cal L}_{0,n}\supset && \eta^{\mu_1\nu_1\cdots \mu_n\nu_n}\partial_{\mu_1}\partial_{\nu_1}\phi \cdots\partial_{\mu_n}\partial_{\nu_n}\phi
=\partial_{\mu_1}\(\eta^{\mu_1\nu_1\cdots \mu_n\nu_n}\partial_{\nu_1}\phi \cdots\partial_{\mu_n}\partial_{\nu_n}\phi\).\eea
The leading term contains one tensor and $n-1$ scalars,
\be \label{0decterms} {\cal L}_{0,n}\supset \eta^{\mu_1\nu_1\cdots \mu_n\nu_n}h_{\mu_1\nu_1}\partial_{\mu_2}\partial_{\nu_2}\phi \cdots\partial_{\mu_n}\partial_{\nu_n}\phi.\ee 
We can write this term as $\sim h_{\mu\nu}X^{\mu\nu}_{(n)}$, where $X_{(n)}^{\mu\nu}\sim\eta^{\mu\nu\mu_2\nu_2\cdots \mu_n\nu_n}\partial_{\mu_2}\partial_{\nu_2}\phi \cdots\partial_{\mu_n}\partial_{\nu_n}\phi$ are the same identically conserved tensors which appear in the decoupling limit of dRGT theory \cite{deRham:2010ik}.  Note that the lowest order term, the tadpole ${\cal L}_{0,1}$, is gauge invariant to begin with and hence generates no St\"uckelberg fields.

 The terms \eqref{0decterms} are invariant up to a total derivative under the decoupling limit linear diffeomorphism symmetry \eqref{decsym}, as can be checked explicitly using integrations by parts and the anti-symmetry of the $\eta$ symbol \eqref{tensor}.  They are ghost free (as are all the decoupling-limit terms we will derive, since they descend from ghost-free terms) and yield second second order equations of motion for both $h$ and $\phi$.

We may also find the contribution of the vector modes to the decoupling limit (the corresponding terms for the full dRGT theory are known only in part \cite{deRham:2010gu,Koyama:2011wx,Tasinato:2012ze}).  The terms with a single vector and $n-1$ scalars are total derivatives.  The leading contribution comes from terms with two powers of $\partial A$ and $n-2$ powers of $\partial\partial\phi$, and can be written as
\be  {\cal L}_{0,n}\supset \eta^{\mu_1\nu_1\cdots \mu_n\nu_n}F_{\mu_1\mu_2}F_{\nu_1\nu_2}\partial_{\mu_3}\partial_{\nu_3}\phi \cdots\partial_{\mu_n}\partial_{\nu_n}\phi,\ee 
where $F_{\mu\nu}\equiv \partial_\mu A_\nu-\partial_\nu A_\mu$ is the usual Maxwell field strength corresponding to $A_\mu$.  
These terms are manifestly gauge invariant under the decoupling limit Maxwell gauge symmetry \eqref{decsym}, and yield second order equations of motion for both $A_\mu$ and $\phi$.  (In fact, they are instances of the mixed $p$-form galileons introduced in \cite{Deffayet:2010zh}.)

\subsection{Two-derivative terms}

We next turn to the St\"uckelberg expansion of the two-derivative terms \eqref{2d}.
The first non-trivial term, the kinetic term ${\cal L}_{2,2}$, is gauge invariant and generates no St\"uckelberg fields.  The rest of the terms ${\cal L}_{2,n}$, $n\geq 3$, give vanishing scalar self-interactions (when all the $h$'s are replaced with $\partial\partial\phi$, there is a $\phi$ with four derivatives that vanishes by the anti-symmetry of the $\eta$ symbol).  In addition, there are no self-interactions with one tensor and $n-1$ scalars, because by the anti-symmetry of $\eta$ symbol, the $h$ which remains must be the one with the derivatives, and this term is a total derivative,
\bea {\cal L}_{2,n}\supset && \eta^{\mu_1\nu_1\cdots \mu_{n+1}\nu_{n+1}}\partial_{\mu_1}\partial_{\nu_1}h_{\mu_2\nu_2}\, \partial_{\mu_3}\partial_{\nu_3}\phi\cdots \partial_{\mu_{n+1}}\partial_{\nu_{n+1}}\phi\nn\\
&& =\partial_{\mu_1}\( \eta^{\mu_1\nu_1\cdots \mu_{n+1}\nu_{n+1}}\partial_{\nu_1}h_{\mu_2\nu_2}\, \partial_{\mu_3}\partial_{\nu_3}\phi\cdots \partial_{\mu_{n+1}}\partial_{\nu_{n+1}}\phi \).
\eea

The leading term that will appear in the decoupling limit contains two powers of $h$ and $n-2$ powers of $\partial\partial\phi$,
\be {\cal L}_{2,n}\supset  \eta^{\mu_1\nu_1\cdots \mu_{n+1}\nu_{n+1}}\partial_{\mu_1}\partial_{\nu_1}h_{\mu_2\nu_2}\, h_{\mu_3\nu_3}\partial_{\mu_4}\partial_{\nu_4}\phi\cdots \partial_{\mu_{n+1}}\partial_{\nu_{n+1}}\phi.
\ee
These terms are invariant under the decoupling limit gauge symmetries up to a total derivative (as can be checked explicitly using the anti-symmetry of the $\eta$ symbol and integration by parts) and yield second order equations of motion for both the scalar and tensor.

In $D=4$, the term ${\cal L}_{2,3}$ gives a new contribution to the scalar-tensor sector of the decoupling limit,
\bea \label{dec42} {\cal L}_{2,3}\supset && \eta^{\mu_1\nu_1\cdots \mu_4\nu_4}\partial_{\mu_1}\partial_{\nu_1}h_{\mu_2\nu_2}\, h_{\mu_3\nu_3}\partial_{\mu_4}\partial_{\nu_4}\phi \nn\\
&&\sim \phi\bigg[R^2-4R_{\mu\nu}R^{\mu\nu}+R_{\mu\nu\rho\sigma}R^{\mu\nu\rho\sigma}\bigg]_{h^2}.
\eea
In the second line we have integrated by parts to take derivatives off of the scalar, and what remains is the ${\cal O}\(h^2\)$ part of the Gauss-Bonnet term.
We conjecture that this term in fact captures the decoupling limit of the as-yet unknown fully non-linear, diffeomorphism non-invariant two-derivative term in $D=4$.  If this is correct, we may add \eqref{dec42} to the decoupling limit of dRGT theory in $D=4$, and its presence may modify the conclusions of studies of massive gravity in the decoupling limit, e.g. \cite{deRham:2010tw,Chkareuli:2011te,Berezhiani:2013dw,Berezhiani:2013dca}.

Turning to the vector modes, no terms with only vectors and scalars survives, because the $\partial\partial h$ piece vanishes upon the St\"uckelberg substitution.  Terms with one tensor (which must be the $\partial\partial h$), one vector and $n-2$ scalars are total derivative.  The leading contribution comes from terms with one tensor, two powers of $\partial A$, and $n-3$ powers of $\partial\partial\phi$, and can be written as
\be {\cal L}_{2,n}\supset  \eta^{\mu_1\nu_1\cdots \mu_{n+1}\nu_{n+1}}\partial_{\mu_1}\partial_{\nu_1}h_{\mu_2\nu_2}\, F_{\mu_3\mu_4}F_{\nu_3\nu_4}\partial_{\mu_5}\partial_{\nu_5}\phi\cdots \partial_{\mu_{n+1}}\partial_{\nu_{n+1}}\phi.
\ee
This term is gauge invariant under the decoupling limit St\"uckelberg gauge symmetries \eqref{decsym}, and yields second order equations of motion for $h_{\mu\nu}$, $A_\mu$ and $\phi$.

\subsection{$d$-derivative terms}

Finally we turn to the general $d$-derivative terms \eqref{dd}.  The lowest order term ${\cal L}_{d,d/2+1}$ is gauge invariant and generates no St\"uckelberg fields.  The rest of the terms ${\cal L}_{d,n}$, $n\geq d/2+2$ give no interactions with scalars and $d/2$ or fewer tensors.  The leading term which can appear in the decoupling limit contains $d/2+1$ powers of $h$ and $n-(d/2+1)$ powers of $\partial\partial\phi$,
\bea  {\cal L}_{d,n}\supset  \eta^{\mu_1\nu_1\cdots \mu_{n+d/2}\nu_{n+d/2}}\partial_{\mu_1}\partial_{\nu_1}h_{\mu_2\nu_2}\cdots\partial_{\mu_{d-1}}\partial_{\nu_{d-1}}h_{\mu_d\nu_d}\, h_{\mu_{d+1}\nu_{d+1}}\partial_{\mu_{d+2}}\partial_{\nu_{d+2}}\phi\cdots \partial_{\mu_{n+d/2}}\partial_{\nu_{n+d/2}}\phi .\nn\\ \label{dscterms} \eea
These terms are invariant under the decoupling limit diffeomorphism symmetries \eqref{decsym} up to a total derivative and yield second order equations of motion for both the scalar and tensor, and may be present in the decoupling limit action for ghost-free massive gravities in $D>4$.

Since the terms \eqref{dscterms} are invariant under the linear diffeomorphism symmetries \eqref{decsym} only up to a total derivative, yet its variations with respect to the fields -- which enter as vertices in the Feynman rules for loops -- are strictly invariant, we can anticipate that these terms will satisfy a non-renormalization theorem just as the zero-derivative terms \eqref{0decterms} do\footnote{This is similar in character to the non-renormalization theorems satisfied by the Galileons \cite{Luty:2003vm,Hinterbichler:2010xn}.} \cite{deRham:2012ew}.

Turning to the vectors, no terms where any of the $\partial\partial h$ pieces are turned into a vector or scalar can survive.  Terms with $d/2$ tensors (which must be the $\partial\partial h$), one vector and $n-d/2-1$ scalars are total derivatives.  The leading contribution comes from terms with $d/2$ tensors, two powers of $\partial A$, and $n-d/2-2$ powers of $\partial\partial\phi$, and can be written as
\bea  && {\cal L}_{d,n}\supset  \eta^{\mu_1\nu_1\cdots \mu_{n+d/2}\nu_{n+d/2}}\partial_{\mu_1}\partial_{\nu_1}h_{\mu_2\nu_2}\cdots\partial_{\mu_{d-1}}\partial_{\nu_{d-1}}h_{\mu_d\nu_d}\, \nn\\ && \times F_{\mu_{d+1}\mu_{d+2}}F_{\nu_{d+1}\nu_{d+2}}\partial_{\mu_{d+3}}\partial_{\nu_{d+3}}\phi\cdots \partial_{\mu_{n+d/2}}\partial_{\nu_{n+d/2}}\phi . \eea
These terms are manifestly gauge invariant under the decoupling limit St\"uckelberg gauge symmetries \eqref{decsym}, and yield second order equations of motion for $h_{\mu\nu}$, $A_\mu$ and $\phi$.

\section{\label{conc}Conclusions}

We have studied pseudo-linear interaction terms for a massive graviton, and have written down the possible ghost-free terms, including those with derivative interactions.  We have conjectured that these terms are in one-to-one correspondence with the possible ghost-free interaction terms for fully non-linear massive gravity, and hence characterize the general solution to the problem of writing down ghost-free interaction terms for the massive graviton.  In particular, in $D=4$ there is one new derivative self-interaction term which is ghost-free and may be added to dRGT theory.

It is likely that the pseudo-linear terms are not phenomenologically viable, since they reduce to linearized gravity in the massless limit rather than fully non-linear GR. That being said, the non-linearities of GR have for the most part not been directly tested, and it may be interesting to check to what extent the pseudo-linear terms are compatible with precision solar system data and binary pulsar data.  

Even apart from questions of phenomenology, these pseudo-linear terms are of purely field-theoretic interest as ghost-free massive gravities which may help us in the study of the fully non-linear theory.  In particular, based on the existence of terms in the pseudo-linear theory, we can conjecture the existence of as yet unknown fully non-linear ghost-free derivative interactions and determine some of their properties, such as the form of the decoupling limit.  
Since the pseudo-linear terms are comparatively simple to work with, they should prove useful as toy models to better understand the features of interacting massive gravitons.

{\bf Acknowledgements:} The author would like to thank Rachel Rosen and Claire Zukowski for comments on the manuscript, and James Bonifacio for correcting a typo in v2.  Research at Perimeter Institute is supported by the Government of Canada through Industry Canada and by the Province of Ontario through the Ministry of Economic Development and Innovation.  This work was made possible in part through the support of a grant from the John Templeton Foundation. The opinions expressed in this publication are those of the author and do not necessarily reflect the views of the John Templeton Foundation. 

\appendix 

\section{\label{screensec}A simple degravitating solution}

 In this Appendix we display a simple degravitating solution (i.e. a solution which is flat $h_{\mu\nu}\sim \eta_{\mu\nu}$ despite the presence of a cosmological term) supported by the pseudo-linear interaction terms.  This will serve to illustrate some ways in which the pseudo-linear interaction terms display features of their fully non-linear counterparts. 

Consider the pseudo-linear theory in $D=4$,
\be {\cal L}={\rm derivative\ terms}\ -2\Lambda M_P h -\frac{1}{2}m^2(h_{\mu\nu}h^{\mu\nu}-h^2)+{m^2\alpha \over 6 M_P} \(h^3-3h h_{\mu\nu}^2+2 h_{\mu\nu}^3\) .\label{screenlag}\ee
We have included the potential terms \eqref{tensor}.  The term linear in $h$ is the pseudo-linear version of the cosmological constant term with cosmological constant $\Lambda$, the quadratic potential is the Fierz-Pauli mass term with mass $m$, and the dimensionless coefficient $\alpha$ governs the cubic term ${\cal L}_{0,3}$ (we have chosen to set the coefficient of ${\cal L}_{0,4}$ to zero for simplicity, the essential conclusions are unchanged if it is restored).

We look for solutions that are proportional to flat space,
\be h_{\mu\nu}=c\,\eta_{\mu\nu},\ee
for some constant $c$.
Plugging this into the equations of motion of \eqref{screenlag}, we find an equation for $c$,
\be 3 \alpha m^2 c^2 +3 m^2 M_P c - 2 M_P^2\Lambda=0,\ee
with solutions
\be c=-{M_P\over 2\alpha}\(1\pm\sqrt{1+{8\over 3}\alpha{\Lambda\over m^2}}\).\label{csolution}\ee
Solutions exist for ${\Lambda\over m^2}\alpha\geq -{3\over 8}$.  In particular, for the case of interest $|\Lambda|\gg m^2$, solutions always exist for $\alpha$ of ${\cal O}(1)$ if the sign of $\alpha$ is chosen to coincide with the sign of $\Lambda$.

If we now expand the Lagrangian \eqref{screenlag} in fluctuations $\tilde h_{\mu\nu}$ about the solution,
\be h_{\mu\nu}=c\,\eta_{\mu\nu}+\tilde h_{\mu\nu},\ee
we find
\be  {\cal L}={\rm derivative\ terms}\ -\frac{1}{2}m^2\sqrt{1+{8\over 3}\alpha{\Lambda\over m^2}}\(\tilde h_{\mu\nu}\tilde h^{\mu\nu}-\tilde h^2\)+{m^2\alpha \over 6 M_P} \(\tilde h^3-3\tilde h \tilde h_{\mu\nu}^2+2\tilde h_{\mu\nu}^3\) .\label{screenlagfluc}\ee

Around the new solution, the mass for the fluctuations is renormalized to the value $\tilde m^2\equiv m^2\sqrt{1+{8\over 3}\alpha{\Lambda\over m^2}}$, and the cutoff scale around the new solution is $\tilde \Lambda_3^3=\(1+{8\over 3}\alpha{\Lambda\over m^2}\)\Lambda_3^3$.  We see that the mass term vanishes when the two branches of solutions \eqref{csolution} coincide, and the cutoff scale drops to zero.  At this special point, the quadratic action has fewer degrees of freedom than the full theory and the solution is infinitely strongly coupled\footnote{This is not to say these solutions are not viable or are somehow inconsistent, their fluctuations are just beyond the reach of the standard perturbation theory.}.  The vanishing of quadratic degrees of freedom also occurs around self-accelerating solutions in the full dRGT theory \cite{Gumrukcuoglu:2011zh}.  

For $\Lambda\sim M_P^2$ and $\alpha\sim 1$, the renormalized mass is $\tilde m^2\sim M_P m$, much too large to be the long-range gravity we know, and the cutoff $\tilde \Lambda_3\sim \({M_P^2\over m^2}\)^{1/3}\Lambda_3\sim M_P$ is very large, so the corresponding Vainshtein radius is much too small to be relevant for the solar system.  These are the same sorts of issues that arise for the degravitating solutions in the full dRGT theory and its decoupling limit \cite{deRham:2010tw,D'Amico:2011jj}.

\bibliographystyle{utphys}
\addcontentsline{toc}{section}{References}
\bibliography{linearmassive}

\end{document}